\pdfoutput=1

\documentclass[11pt,twoside,a4paper,cmspaper,final,collab]{cms-tdr}
\def\svnVersion{491044P}\def\cmsCernNoTag{CERN-EP-2018-256}\def\cmsCernDate{\today}\def\cmsMessage{Published in Physics Letters B as \href{http://dx.doi.org/10.1016/j.physletb.2019.02.018}{\doi{10.1016/j.physletb.2019.02.018.}}}

\begin{document}\cmsNoteHeader{HIN-18-010}

\hyphenation{had-ron-i-za-tion}
\hyphenation{cal-or-i-me-ter}
\hyphenation{de-vices}
\RCS$HeadURL: svn+ssh://svn.cern.ch/reps/tdr2/papers/HIN-18-010/trunk/HIN-18-010.tex $
\RCS$Id: HIN-18-010.tex 489086 2019-02-14 16:40:41Z alverson $

\newlength\cmsFigWidth
\ifthenelse{\boolean{cms@external}}{\setlength\cmsFigWidth{0.49\textwidth}}{\setlength\cmsFigWidth{0.65\textwidth}}
\ifthenelse{\boolean{cms@external}}{\providecommand{\cmsLeft}{upper\xspace}}{\providecommand{\cmsLeft}{left\xspace}}
\ifthenelse{\boolean{cms@external}}{\providecommand{\cmsRight}{lower\xspace}}{\providecommand{\cmsRight}{right\xspace}}

\newcommand{\ctauxyz}{\ensuremath{\ell_{\cPJgy}^{\mathrm{3D}}}\xspace}
\newcommand{\Lxyz}{\ensuremath{L_{xyz}}\xspace}
\newcommand {\deta}     {\ensuremath{\Delta\eta}}
\newcommand {\dphi}     {\ensuremath{\Delta\phi}}
\newcommand{\ket}       {\ensuremath{KE_{\mathrm{T}}}\xspace}
\newcommand {\PbPb}  {\ensuremath{\text{PbPb}}\xspace}
\newcommand {\AuAu}  {\ensuremath{\text{AuAu}}\xspace}
\newcommand {\pPb}  {\ensuremath{\Pp\text{Pb}}\xspace}
\newcommand {\AonA}  {\ensuremath{\text{AA}}\xspace}
\newcommand {\pA}    {\ensuremath{\Pp\text{A}}\xspace}
\newcommand{\noff}    {\ensuremath{N_\mathrm{trk}^\mathrm{offline}}\xspace}
\newcommand {\roots}    {\ensuremath{\sqrt{s}}}

\cmsNoteHeader{HIN-18-010}
\title{Observation of prompt \cPJgy\ meson elliptic flow in high-multiplicity \pPb collisions at \texorpdfstring{$\sqrtsNN = 8.16$\TeV}{sqrt{s[NN]} = 8.16 TeV}}

\date{\today}

\abstract{
A measurement of the elliptic flow ($v_2$) of prompt \cPJgy\ mesons in high-multiplicity \pPb\ collisions
is reported using data collected by the CMS experiment
at a nucleon-nucleon center-of-mass energy $\sqrtsNN = 8.16\TeV$.
Prompt \cPJgy\ mesons decaying into two muons are reconstructed in the
rapidity region in the nucleon-nucleon center-of-mass frame ($y_\mathrm{cm}$),
corresponding to either $-2.86<y_\mathrm{cm}<-1.86$ or $0.94<y_\mathrm{cm}<1.94$.
The average $v_2$ result from the two rapidity ranges is reported over the
transverse momentum (\pt) range from 0.2 to 10\GeV. Positive $v_2$ values are
observed for the prompt \cPJgy\ meson, as extracted from long-range
two-particle correlations with charged hadrons, for $2<\pt<8$\GeV.
The prompt \cPJgy\ results are compared with previous CMS measurements of
elliptic flow for open charm mesons (\PDz) and strange hadrons. From these
measurements, constraints can be obtained on the collective dynamics of charm
quarks produced in high-multiplicity events arising from small systems.
}

\hypersetup{
pdfauthor={CMS Collaboration},
pdftitle={Observation of prompt J/psi meson elliptic flow in high-multiplicity pPb collisions at sqrt{s[NN]} = 8.16 TeV},
pdfsubject={CMS},
pdfkeywords={CMS, physics, heavy ion, correlation, flow, pPb, heavy flavor}}

\maketitle

\section{Introduction}

Strong collective behavior is found in the azimuthal correlations of particles
emitted in relativistic nucleus-nucleus (\AonA) collisions
at the BNL RHIC~\cite{Alver:2008gk,Adams:2005ph,Abelev:2009af,Alver:2008gk,Alver:2009id}
and at the CERN LHC~\cite{Chatrchyan:2011eka,Chatrchyan:2012wg,Aamodt:2010pa,ATLAS:2012at,Chatrchyan:2012ta,CMS:2013bza}.
These correlations, which are long-range in pseudorapidity ($\eta$), suggest the
formation of a strongly interacting quark-gluon plasma (QGP) that exhibits
nearly ideal hydrodynamic behavior~\cite{Ollitrault:1992bk,Heinz:2013th,Gale:2013da}.
The azimuthal correlation structure of emitted particles is
typically characterized by its Fourier components~\cite{Voloshin:1994mz}. In particular,
within a hydrodynamic picture, the second and third Fourier anisotropy components
are known as elliptic ($v_2$) and triangular ($v_3$) flow, respectively, and
reflect the QGP medium response to the initial collision geometry and
its fluctuations~\cite{Alver:2010dn,Schenke:2010rr,Qiu:2011hf}.
In recent years, similar long-range collective azimuthal correlations
have also been observed in events with high final-state particle multiplicity in proton-proton
(\Pp\Pp)~\cite{Khachatryan:2010gv,Aad:2015gqa,Khachatryan:2015lva,Khachatryan:2016txc}, proton-nucleus (\pA)~\cite{CMS:2012qk,alice:2012qe,Aad:2012gla,Aaij:2015qcq,ABELEV:2013wsa,Khachatryan:2014jra,Khachatryan:2015waa,Aaboud:2017acw,Aaboud:2017blb},
and lighter \AonA\ collisions~\cite{Adamczyk:2015xjc,Adare:2015ctn,Aidala:2017ajz},
raising the question of whether a fluid-like QGP is created in these much smaller systems.
While experimental measurements in these small systems are consistent with
the hydrodynamic expansion of a tiny QGP droplet, alternative scenarios
based on gluon saturation in the initial state also claim to
capture the main features of the correlation data (recent reviews are provided in Refs.~\cite{Dusling:2015gta,Nagle:2018nvi}).

Because of their large masses, heavy quarks (charm and bottom) are primarily
produced via hard-scattering processes at a very early stage of the collision.
Thus, they are largely decoupled from the bulk production of soft gluons and light-flavor quarks at a later stage
in \AonA\ collisions, and thereby probe the properties and dynamics of the QGP through
its entire evolution~\cite{BraunMunzinger:2007tn}. A strong elliptic flow ($v_2$) signal has been observed for
open heavy-flavor \PDz\ mesons in both \AuAu\ collisions at RHIC~\cite{Adamczyk:2017xur}
and \PbPb\ collisions at the LHC~\cite{Abelev:2014ipa,Acharya:2017qps,Sirunyan:2017plt},
suggesting that charm quarks may develop strong collective flow behavior.
Furthermore, a recent measurement of the elliptic flow of \cPJgy\ mesons in PbPb collisions
at $\sqrtsNN\ = 5.02\TeV$ \cite{Acharya:2017tgv} has provided additional evidence for the collective
behavior of charm quarks in the QGP.

In the study of collectivity in small systems, such as that occurring in \Pp\Pp\ or \pPb\ collisions,
a key open question is whether the strong collective behavior observed for bulk constituents
in high-multiplicity events also extends to charm and bottom quarks.
Long-range correlations involving inclusive muons at high transverse
momentum (\pt) reveal a hint of heavy-flavor quark collectivity in \pPb\ collisions~\cite{Adam:2015bka}.
Furthermore, the recent observation of a significant elliptic flow signal
for prompt \PDz\ mesons in \pPb\ collisions has provided evidence for
charm quark collectivity in a small system~\cite{Sirunyan:2018toe}. The $v_2$ signal for \PDz\ mesons
is found to be smaller than that of light-flavor mesons at a given \pt,
indicating that in these small systems there is a weaker collective motion
for charm quarks, as compared to that of the bulk medium.
However, as the \PDz\ meson carries both a light and a charm quark, the relative contribution of
these different flavor quarks to
the observed $v_2$ signal is not fully constrained. Without detailed theoretical modeling,
a scenario is not excluded where the \PDz\ meson $v_2$ signal is entirely carried by the light-flavor quark.
The observation of an elliptic flow signal for \cPJgy\ mesons in a small system could provide more direct
evidence of charm quark collectivity and could impose new constraints on the collective
dynamics of heavy-quark production in such collisions.
Furthermore, heavy-quark collectivity may also provide a hint of
how, in small systems, hard probes interact with the QGP~\cite{BraunMunzinger:2007tn},
assuming this is formed. First measurement of inclusive \cPJgy\
(combined charmonia and \cPJgy\ mesons from decay of open beauty hadrons) $v_2$ in \pPb\ collisions
was reported in Ref.~\cite{Acharya:2017tfn}, where positive $v_2$ coefficients
were found in the range of $3<\pt<6\GeV$ with center-of-mass rapidities $-4.46<y_\mathrm{cm}<-2.96$ or $2.03<y_\mathrm{cm}<3.53$.
A recent model calculation of \cPJgy\ $v_2$ in \pPb\ collisions suggests little $v_2$ signal
arising from final-state interactions between charm quarks and the QGP medium~\cite{Du:2018wsj}.

This Letter presents the first measurement of prompt \cPJgy\
meson elliptic flow (excluding contributions from \PQb\ hadron decays) from long-range
two-particle correlations in very high multiplicity \pPb\ collisions at $\sqrtsNN\ = 8.16\TeV$.
The $v_2$ harmonics for prompt \cPJgy\ mesons in the ranges
$-2.86<y_\mathrm{cm}<-1.86$ and $0.94<y_\mathrm{cm}<1.94$
are determined over a wide \pt\ range from 0.2 to 10\GeV. To estimate
the possible residual contribution from back-to-back jet-like correlations, the $v_2$
values are also presented after subtracting correlations obtained from
low-multiplicity \pPb\ events (denoted as $v_{2}^\text{sub}$), where jet-like correlations are assumed to dominate.
The results are compared to those of the light strange-flavor \PKzS\ and \PgL\ hadrons, and
the open heavy-flavor prompt \PDz\ meson,
which were previously reported by CMS~\cite{Sirunyan:2018toe} in the same \pt\ range but
in a different rapidity range of $-1.46 < y_{\text{cm}} < 0.54$.
In order to explore possible collectivity at the partonic level,
a comparison is also presented in terms of the transverse kinetic
energy per constituent quark ($\ket$/$n_\Pq$,
where $\ket = \sqrt{\smash[b]{m^2 + \pt^2}} - m$, and $n_\Pq$ is
the number of constituent quarks).

\section{The CMS detector}

The central feature of the CMS apparatus is a superconducting solenoid of 6\unit{m}
internal diameter, providing a magnetic field of 3.8\unit{T}. Within the solenoid volume,
there are four primary subdetectors including a silicon
pixel and strip tracker detector, a lead tungstate crystal electromagnetic calorimeter,
and a brass and scintillator hadron calorimeter, each composed of a
barrel and two endcap sections.  Iron and quartz-fiber Cherenkov hadron forward
(HF) calorimeters cover the range $3.0 < \abs{\eta} < 5.2$. Muons are measured in the range
$\abs{\eta} < 2.4$ in gas-ionization detectors embedded in the steel flux-return
yoke outside the solenoid, with detection planes made using three technologies:
drift tubes, cathode strip chambers, and resistive-plate chambers. The silicon tracker
measures charged particles within the range $\abs{\eta}< 2.5$. For charged
particles with $1 < \pt < 10\GeV$ and $\abs{\eta} < 1.4$, the track resolutions are
typically 1.5\% in \pt and 25--90 (45--150)\mum in the transverse (longitudinal)
impact parameter~\cite{Chatrchyan:2014fea}. A detailed description of the CMS detector,
together with a definition of the coordinate system used and the relevant kinematic variables,
can be found in Ref.~\cite{Chatrchyan:2008zzk}.

\section{Data selection and \texorpdfstring{\cPJgy}{J/psi} meson reconstruction}

The \pPb\ data at $\sqrtsNN\ = 8.16\TeV$ used in this analysis were
collected in 2016, and correspond to an integrated luminosity
of 186\nbinv. The beam energies are 6.5\TeV for the protons and
2.56\TeV per nucleon for the lead nuclei. Because of the asymmetric beam conditions,
particles selected in the laboratory rapidity range of $1.4<y_\text{lab}<2.4$
($-2.4<y_\text{lab}<-1.4$) have a corresponding nucleon-nucleon center-of-mass
frame rapidity range of $0.94<y_\mathrm{cm}<1.94$ ($-2.86<y_\mathrm{cm}<-1.86$),
with positive rapidity defined in the proton beam direction.
To minimize statistical uncertainties, the quoted \cPJgy\ meson $v_2$ results
combine the individual values obtained for the proton and lead beam directions.

The \pPb\ data are analyzed in different ranges of \noff, where \noff\ is the number of
primary charged particle tracks~\cite{Chatrchyan:2014fea} with $\abs{\eta}<2.4$ and $\pt > 0.4\GeV$.
The main results are obtained with events in the high-multiplicity range $185 \leq \noff < 250$.
To select these events, dedicated triggers were developed, as discussed
in Refs.~\cite{Sirunyan:2017quh,Sirunyan:2017uyl}.
Events with $\noff < 35$ are also used to estimate the possible contribution of residual
back-to-back jet-like correlations. These lower-multiplicity events are selected online with a
hardware-based trigger requiring two muon candidates in the muon detectors
with no explicit momentum or rapidity threshold~\cite{Chatrchyan:2013nza}.
In the offline analysis, hadronic collisions are selected by requiring at
least one HF calorimeter tower with more than 3\GeV of total energy
in each of the two HF detectors. Events must contain a primary vertex
close to the nominal interaction point of the beams, within 15\unit{cm}
along the beam direction, and 0.2\unit{cm} in the plane transverse to beam direction.
The \noff\ range limits correspond to fractional inelastic cross sections
from 100 to 57\% for $\noff<35$, and from 0.33 to 0.01\% for $185 \leq \noff < 250$, respectively.

The offline muon reconstruction algorithm starts either by finding tracks in the muon detectors,
which are then fitted together with tracks reconstructed in the silicon tracker (\emph{global muons}),
or by extrapolating tracks from the silicon tracker to match a hit on at least one segment of
the muon detectors (\emph{tracker muons}).
The muon candidates are required to pass the identification criteria of the particle-flow algorithm~\cite{Sirunyan:2017ulk},
which suppresses contamination of ``punch-through'' hadrons misidentified as muons, based on
energy deposition in the calorimeters. The \emph{soft}
muon selection criteria are also imposed, as defined in Ref.~\cite{Chatrchyan:2012xi}, to
further improve the purity of muons.

The \cPJgy\ meson candidates are formed from pairs of oppositely charged muons, originating from a
common vertex.
Based on the vertex probability distributions for signal and background candidates,
the probability that the dimuon pair shares a common vertex is required to be
larger than 1\%, lowering the background from random combinations as well as from semileptonic decays of bottom and charm hadrons.
Because of the long lifetime of \PQb hadrons compared to that of \cPJgy\ mesons,
the nonprompt \cPJgy\ meson component can be reduced by placing constraints on the pseudo-proper decay length~\cite{Buskulic:1993aa}.
This is defined by
\begin{linenomath}
\begin{equation}
\ctauxyz = \Lxyz \frac{m_{\cPJgy}}{\abs{p_{\mu\mu}}},
\end{equation}
\end{linenomath}
where $\Lxyz$ is the distance between the primary and dimuon vertices, $m_{\cPJgy}$ is the Particle Data
Group \cite{pdg} world average value of the \cPJgy\ meson mass (assumed for all dimuon candidates), and
$p_{\mu\mu}$ is the dimuon momentum.
The upper limit (decreasing as a function of \pt) imposed on the $\ctauxyz$
value is based on Monte Carlo (MC) studies with simulated
event samples of \PYTHIA 8.209~\cite{Sjostrand:2007gs,Khachatryan:2015pea},
and found to reject 75--90\% (from low to high \pt) of nonprompt \cPJgy\ mesons, largely independent of
multiplicity. The residual nonprompt \cPJgy\ meson
fraction in the data is estimated to be approximately 5\% across the full \pt\ range, and its effect
on the $v_2$ measurement is propagated as a systematic uncertainty, as described in Section~\ref{sec:syst}.

\section{Analysis technique}

The azimuthal anisotropy of \cPJgy\ mesons
is extracted from the long-range ($\abs{\deta}>1$) two-particle azimuthal correlations,
following an identical procedure to that described in
Refs.~\cite{Khachatryan:2016txc,Khachatryan:2014jra,Sirunyan:2018toe}.
A two-dimensional (2D) correlation function is constructed by pairing each \cPJgy\ candidate
with reference primary charged-particle tracks with $0.3<\pt<3$\GeV and $\abs{\eta}<2.4$
(denoted as ``ref'' particles), and calculating
\begin{linenomath}
\begin{equation}
\label{eq:2Dcorr}
\frac{1}{N_\text{\cPJgy}}\frac{\rd^{2}N^\text{pair}}{\rd\Delta\eta\, \rd\Delta\phi}
= B(0,0)\frac{S(\Delta\eta,\Delta\phi)}{B(\Delta\eta,\Delta\phi)},
\end{equation}
\end{linenomath}
where $\Delta\eta$ and $\Delta\phi$ are the differences in $\eta$
and in the azimuthal angle ($\phi$) of the pair. The same-event pair distribution, $S(\Delta\eta,\Delta\phi)$,
represents the yield of particle pairs normalized by the number of \cPJgy\
candidates from the same event. The mixed-event pair yield distribution,
$B(\Delta\eta,\Delta\phi)$, is constructed by pairing \cPJgy\ candidates in each
event with the reference primary charged-particle tracks from 20 different randomly selected
events, from the same \noff\ range and having a primary vertex falling in the same
2\unit{cm} wide range of reconstructed longitudinal, $z$ coordinate.
The analysis procedure is performed in each \pt\ and invariant mass ($m_\text{inv}$) range of \cPJgy\ candidates.
A correction for the acceptance and efficiency of the \cPJgy\ meson yields
is applied, but found to have a negligible effect on the measurements.
The \dphi\ correlation functions averaged over $\abs{\deta}>1$ (to
remove short-range correlations, such as jet fragmentation) are then obtained from
the 2D distributions and fitted by the first three terms of a Fourier series (including additional terms
has a negligible effect on the fit results):
\begin{linenomath}
\begin{equation}
\label{eq:Vn}
\frac{1}{N_\text{\cPJgy}}\frac{\rd N^\text{pair}}{\rd\Delta\phi} = \frac{N_{\text{assoc}}}{2\pi} \left[ 1+\sum_{n=1}^{3} 2V_{n\Delta} \cos (n\Delta\phi)\right].
\end{equation}
\end{linenomath}
Here, $V_{n\Delta}$ are the Fourier coefficients and $N_{\text{assoc}}$
represents the total number of same-event pairs per \cPJgy\ candidate for a given
invariant mass interval. By assuming that $V_{n\Delta}$ is the product of
single-particle anisotropies of \cPJgy\ mesons and reference charged particles~\cite{Chatrchyan:2013nka},
$V_{n\Delta}(\cPJgy,\text{ref})=v_{n}(\cPJgy) \times v_{n}(\text{ref})$,
the $v_n$ anisotropy harmonics for \cPJgy\ candidates can be extracted as a function of
invariant mass, $v_{n}(\cPJgy) = V_{n\Delta}(\cPJgy,\text{ref})/\sqrt{V_{n\Delta}(\text{ref},\text{ref})}$.
\noindent The $V_{n\Delta}(\text{ref},\text{ref})$ represents the Fourier coefficients extracted by
correlating two reference charged particles. With the current data, only the second order ($n=2$) elliptic
anisotropy harmonic can be measured with meaningful statistical precision.

\begin{figure*}[thb]
\centering
\includegraphics[width=0.9\textwidth]{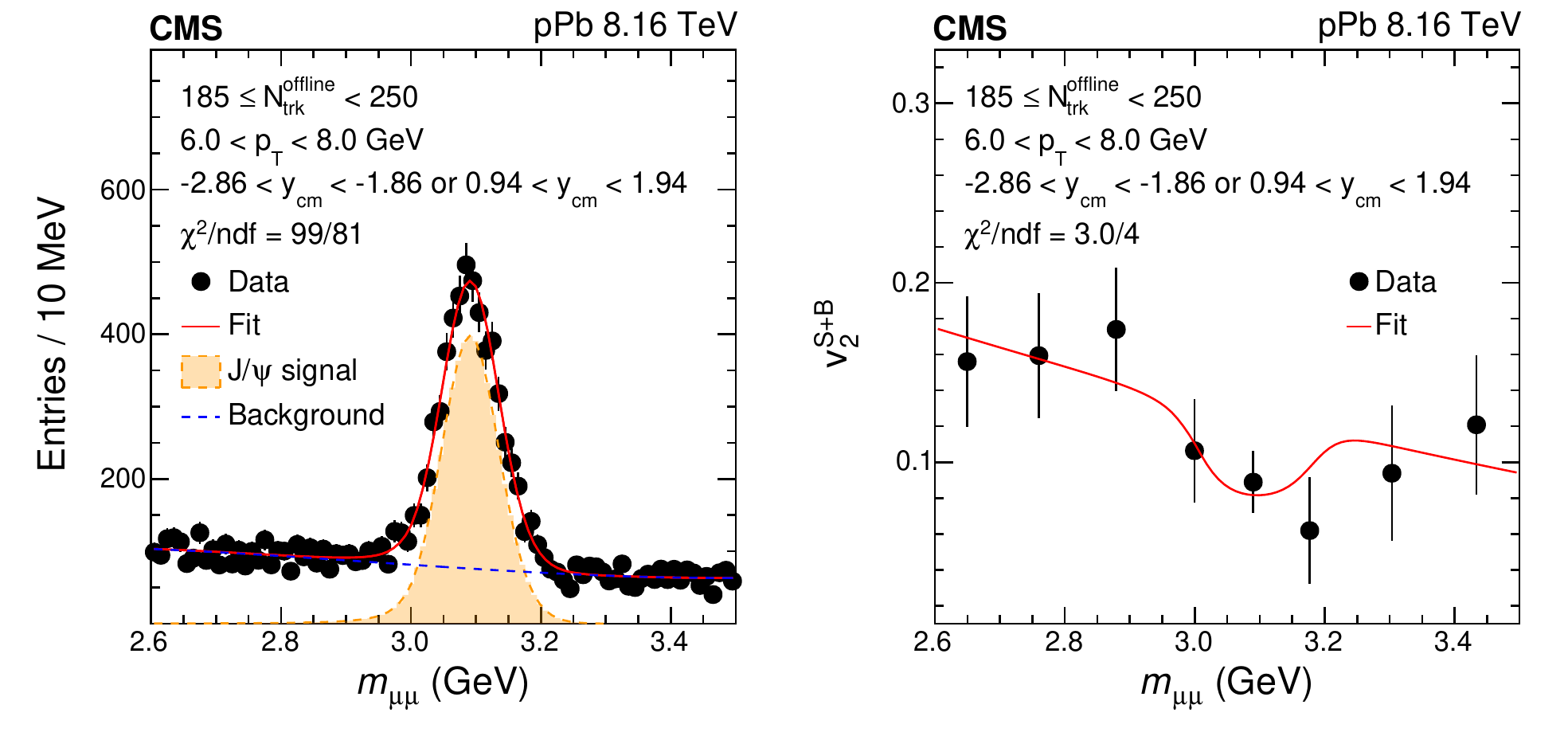}
\caption{Example of fits to the invariant mass spectrum (left)
and the $v_{2}^{S+B}(m_\text{inv})$ distribution (right)
in the \pt interval 6.0--8.0\GeV for \pPb events with $185 \leq \noff < 250$.}
\label{fig:mass_vn_fit}
\end{figure*}

To extract the genuine $v_{2}$ values of the \cPJgy\ meson signal ($v_{2}^{S}$),
the contribution from background candidates ($v_{2}^{B}$) has to be subtracted from
the $v_2$ values of all \cPJgy\ meson candidates, as obtained in the previous step.
The procedure is to first fit the dimuon mass spectrum with a function composed of three components:
two Crystal Ball functions~\cite{SLAC-R-236}
with different widths but common mean and tail parameters for the \cPJgy\ signal (the tail parameters are fixed to
the values obtained from simulation), $S(m_\text{inv})$, and an exponential function
to model the background, $B(m_\text{inv})$. Then,
the signal plus background $v_{2}^{S+B}(m_\text{inv})$ distribution is fitted with:
\begin{linenomath}
\begin{equation}
v_{2}^{S+B}(m_\text{inv}) = \alpha(m_\text{inv})v_{2}^{S} + [1-\alpha(m_\text{inv})]v_{2}^{B}(m_\text{inv}),
\end{equation}
\end{linenomath}
where
\begin{linenomath}
\begin{equation}
\alpha(m_\text{inv}) = \frac{S(m_\text{inv})}{S(m_\text{inv}) + B(m_\text{inv})}.
\end{equation}
\end{linenomath}
Here, $v_{2}^{B}(m_\text{inv})$ for the background \cPJgy\ candidates is modeled as an
exponential function of the invariant mass, and $\alpha(m_\text{inv})$ is the \cPJgy\ signal
fraction obtained from the mass spectrum fit. An example of fits to the mass
spectrum and $v_{2}^{S+B}(m_\text{inv})$ in the \pt\ interval 6.0--8.0\GeV for the
multiplicity range $185 \leq \noff < 250$ is shown in Fig.~\ref{fig:mass_vn_fit}.
The residual contribution of back-to-back dijets to the measured $v_2$ results is estimated
from low-multiplicity \pPb\ events and is removed from the signal after accounting for the
jet yield ratio of the selected events, following a jet subtraction procedure similar to that established in
Refs.~\cite{Chatrchyan:2013nka,Khachatryan:2016txc,Sirunyan:2018toe}.
The Fourier coefficients, $V_{n\Delta}$, extracted from Eq.~(\ref{eq:Vn}) for $\noff < 35$,
are subtracted from the $V_{n\Delta}$ coefficients obtained in the high-multiplicity region, with
\ifthenelse{\boolean{cms@external}}{
\begin{linenomath}
\begin{equation}
\label{eq:vnsubperiph}
\begin{split}
V^\text{sub}_{n\Delta}=&V_{n\Delta}-V_{n\Delta}(\noff<35)\\
&\times\frac{N_\text{assoc}(\noff<35)}{N_\text{assoc}}\\
&\times\frac{Y_\text{jet}}{Y_\text{jet}(\noff<35)}.
\end{split}
\end{equation}
\end{linenomath}
}{
\begin{linenomath}
\begin{equation}
\label{eq:vnsubperiph}
V^\text{sub}_{n\Delta}=V_{n\Delta}-V_{n\Delta}(\noff<35)\,\frac{N_\text{assoc}(\noff<35)}{N_\text{assoc}}\,\frac{Y_\text{jet}}{Y_\text{jet}(\noff<35)}.
\end{equation}
\end{linenomath}
}
Here, $Y_\text{jet}$ represents the jet yield obtained by integrating
the difference of the short-range ($\abs{\deta}<1$) and long-range
event-normalized associated yields for each multiplicity class.
The ratio, $Y_\text{jet}$/$Y_\text{jet}(\noff<35)$, is introduced to account
for the enhanced jet correlations resulting from the selection of higher-multiplicity
events. For $\pt(\cPJgy)<4.5\GeV$, the jet yield ratio cannot be directly estimated
from the two-particle azimuthal correlations, as the \cPJgy\ candidates tend to have larger
$\eta$ values than the acceptance for charged particles.
Therefore, the value is assumed to be the same as that for the high-\pt\
region, where no \pt\ dependence has been observed.
It was also previously observed that the values
of jet yield ratio for \PDz\ and strange particle species show little
dependence on \pt\ over the full \pt\ range~\cite{Sirunyan:2018toe}.

\section{Systematic uncertainties}
\label{sec:syst}

Sources of systematic uncertainties on the prompt \cPJgy\ meson $v_2$ measurement include
the \cPJgy\ meson yield correction (acceptance and efficiency correction derived from \PYTHIA simulation),
the nonprompt \cPJgy\ meson contamination,
the background $v_{2}^{B}(m_\text{inv})$ functional form,
the signal and background invariant mass PDF,
the jet subtraction procedure,
the contamination of events containing more than one \pPb\ interaction (pileup),
and the trigger bias.
In this Letter, the quoted uncertainties in $v_2$ are absolute values,
and are found to have no dependence on \pt, except those for the jet subtraction procedure.
Systematic uncertainties originating from different sources are added in quadrature to obtain
the overall systematic uncertainty shown as boxes in the figures.

To evaluate the uncertainties arising from the efficiency correction to the \cPJgy\ meson yield,
the $v_2$ values are compared to the uncorrected ones, yielding an uncertainty of 0.008.
The effect on the measured $v_2$ due to the residual contribution from
nonprompt \cPJgy\ mesons is evaluated by varying the $\ctauxyz$
requirement such that the nonprompt \cPJgy\ meson yield is doubled.
The $v_2$ values are found not to change by more than $\pm 0.004$,
which is assigned as the systematic uncertainty due to the \cPJgy\
meson yield correction.
Possible differences in the rejection efficiency of nonprompt \cPJgy\ mesons
between data and simulation are investigated and found to be negligible.
The systematic uncertainties from the background $v_{2}$ functional form
are evaluated by comparing $v_{2}^{B}(m_\text{inv})$ values
based on first-, second-, and third-order polynomial fits to the background distribution.
The resulting \cPJgy\ signal $v_2$ values are found to vary by less than 0.009.
Systematic effects related to signal invariant mass PDF are found to be negligible by releasing, one at a time,
the fixed tail parameters of the Crystal Ball functions.
The variation of $v_2$, while changing the background invariant mass PDF to a
second- or third-order polynomial function is also found to be negligible.
In the jet subtraction procedure, the statistical precision of the jet yield ratio is limited.
The $v_{2}^\text{sub}$ results are found to be consistent within
$\pm 0.002$ to $\pm 0.014$ (increasing with \pt) when varying the jet yield ratio by its statistical uncertainty.
The systematic uncertainties from the potential pileup effect and the trigger bias
are taken to be the same as for inclusive charged particles in Ref.~\cite{Sirunyan:2017uyl},
where they can be established with good statistical precision. The pileup and trigger bias
uncertainties are negligible compared to the other sources of systematic uncertainties,
as the fraction of residual pileup events is only a few \% and the trigger efficiency
is close to 100\%.

\section{Results}

\begin{figure}[thb]
\centering
\includegraphics[width=0.48\textwidth]{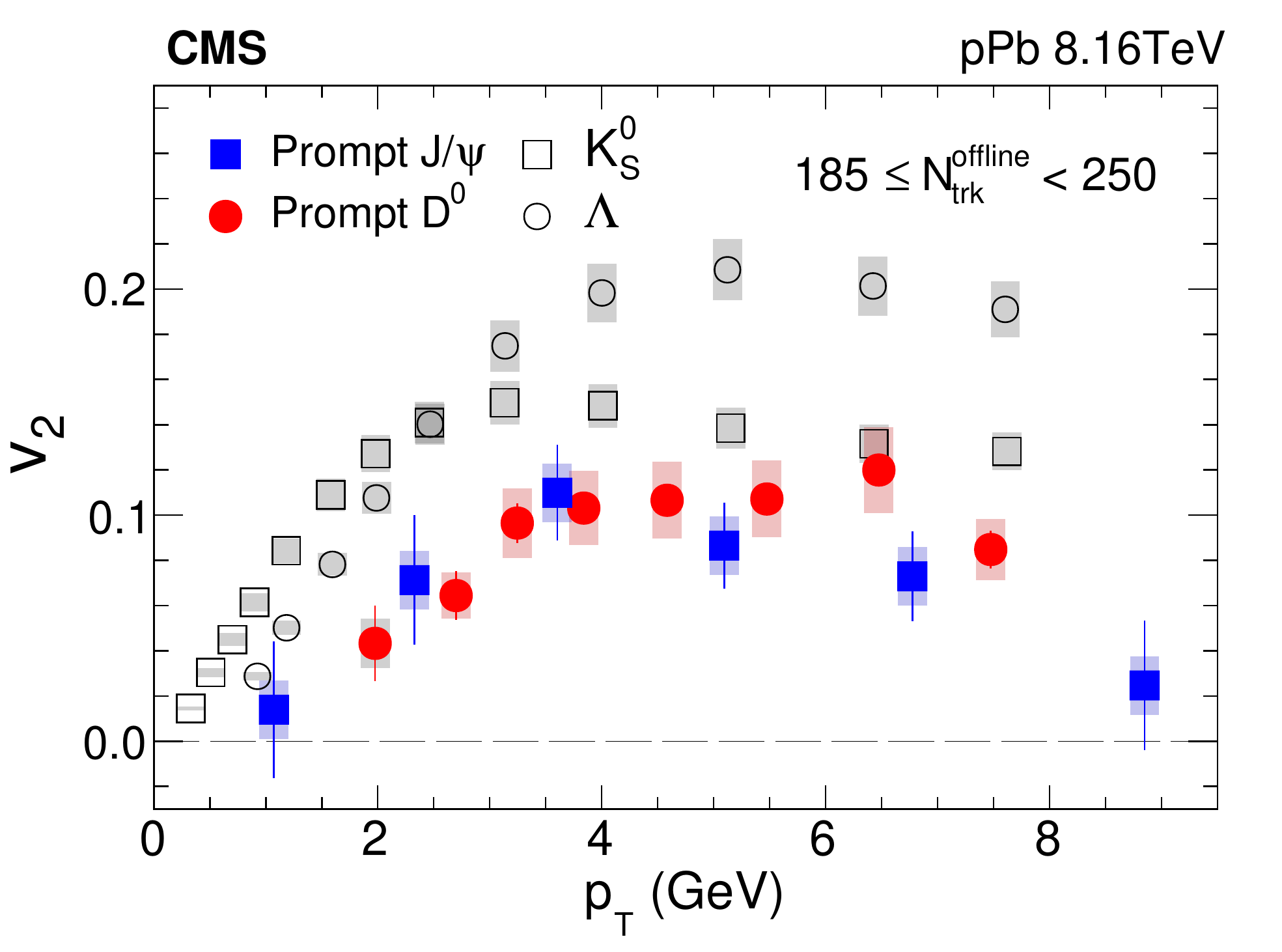}
\caption{The $v_2$ results of the prompt \cPJgy\ mesons at forward rapidities
($-2.86<y_\mathrm{cm}<-1.86$ or $0.94<y_\mathrm{cm}<1.94$), as a function of
\pt\ in the multiplicity range  $185 \leq \noff < 250$ for \pPb\ collisions
at $\sqrtsNN\ = 8.16\TeV$. Data for \PKzS\ and \PgL\ hadrons,
and prompt \PDz\ mesons at midrapidity ($-1.46<y_\mathrm{cm}<0.54$) from
previous CMS measurements~\cite{Sirunyan:2018toe} are also shown for comparison.
The error bars correspond to statistical uncertainties, while the shaded areas denote
the systematic uncertainties.
 }
\label{fig:jpsiv2result}
\end{figure}

Figure~\ref{fig:jpsiv2result} shows the $v_2$ results of prompt \cPJgy\ mesons
at forward rapidities ($-2.86<y_\mathrm{cm}<-1.86$ or $0.94<y_\mathrm{cm}<1.94$)
for high-multiplicity ($185 \leq \noff < 250$) \pPb\ collisions, covering a \pt\
range from 0.2 to 10\GeV.
Results obtained separately for \cPJgy\ meson rapidity in the Pb- and p-going direction
are compared, and found to be consistent within statistical uncertainties.
Thus, as mentioned earlier, combined $v_2$ values are presented for the best statistical precision.
The $v_2$ results for \PKzS\ and \PgL\ hadrons (light, strange-flavor), and prompt \PDz\ mesons (open heavy-flavor),
reported in a previous CMS publication~\cite{Sirunyan:2018toe} for the
midrapidity region $-1.46<y_\mathrm{cm}<0.54$, are also shown for comparison.

Positive prompt \cPJgy\ meson $v_2$ values are observed
over a wide \pt range from about 2 to 8\GeV. The prompt \cPJgy\ meson $v_{2}$
results show a trend of first increasing up to $\pt \sim 4$\GeV and then decreasing
toward higher \pt. This observed trend appears to be in common with the other hadron
species shown.
In the \pt range below 5\GeV, the $v_2$ values for \cPJgy\ and \PDz\ mesons
are consistent with each other within the uncertainties, while an indication of smaller $v_2$
values for \cPJgy\ mesons than that for \PDz\ mesons is seen for $\pt>5$\GeV, although the difference
is not significant within current experimental uncertainties.
Over the full \pt range, the $v_2$ signal values for both \cPJgy\ and \PDz\ hadrons are
smaller than those for \PKzS\ and \PgL\ hadrons. This observation is consistent with
the earlier conclusion that charm quarks develop a weaker collective dynamics
than light quarks in small systems~\cite{Sirunyan:2018toe}.
Because of experimental limitation, $v_2$ values for
the prompt \cPJgy\ meson and the other meson species are not compared within
the same rapidity range, possibly affecting their comparison.
The rapidity dependence of $v_2$ values for charged particles
in \pPb\ collisions has been measured~\cite{Khachatryan:2016ibd,Sirunyan:2017igb},
suggesting up to around 15\% variation from $\abs{y_\text{lab}} \sim 0$ to $2.4$.

\begin{figure}[thb]
\centering
\includegraphics[width=0.48\textwidth]{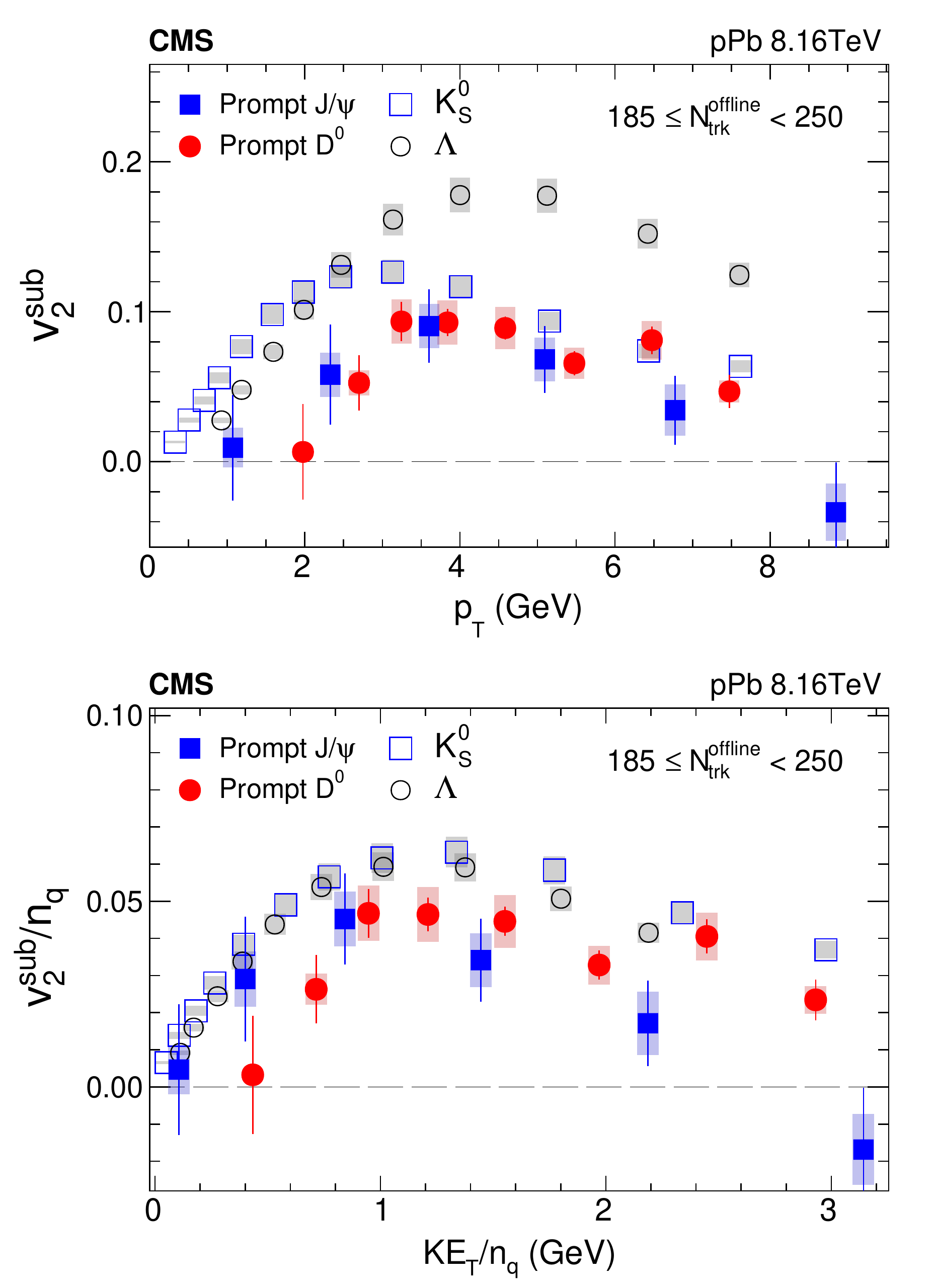}
\caption{Upper: the $v_{2}^\text{sub}$ values for prompt \cPJgy\ mesons
at forward rapidities ($-2.86<y_\mathrm{cm}<-1.86$ or $0.94<y_\mathrm{cm}<1.94$),
as well as for \PKzS\ and \PgL\ hadrons, and prompt \PDz\ mesons at midrapidity ($-1.46<y_\mathrm{cm}<0.54$),
as a function of \pt\ for \pPb\ collisions at $\sqrtsNN\ = 8.16\TeV$
with $185 \leq \noff < 250$. Lower: the $n_\Pq$-normalized $v_{2}^\text{sub}$ results.
The \PKzS, \PgL\ and \PDz\ $v_{2}^\text{sub}$ data are taken from Ref.~\cite{Sirunyan:2018toe}.
The error bars correspond to statistical uncertainties, while the shaded areas denote
the systematic uncertainties.
 }
\label{fig:jpsiv2resultncq}
\end{figure}

To better study the elliptic flow signal coming purely from long-range
collective correlations, the \cPJgy\ $v_2$ results are corrected for residual jet correlations.
The resulting ($v_{2}^\text{sub}$) values are shown in Fig.~\ref{fig:jpsiv2resultncq}
(upper) for prompt \cPJgy\ mesons
as a function of \pt\ with $185 \leq \noff < 250$, and compared to similarly
corrected \PKzS, \PgL, and \PDz\ hadron results~\cite{Sirunyan:2018toe}.
The effect of the correction for all particle species is most noticeable at very high \pt,
while the overall \pt\ dependence of the $v_2$ data
remains unchanged. The \PKzS\ mesons have a larger correction applied to their $v_2$ values
(possibly because \PKzS\ mesons are more correlated with the bulk multiplicity, and thus are biased toward
stronger jet correlations due to the selection of high multiplicities)
and their $v_{2}^\text{sub}$ values after the correction tend to converge
to those of the prompt \cPJgy\ and \PDz\ mesons at high \pt.

A recent model calculation of \cPJgy\ $v_2$ in minimum bias \pPb\ collisions, based on final-state interactions
between produced charm quarks and a QGP medium, suggests a very small $v_2$ signal of less than
0.01~\cite{Du:2018wsj}. This calculation indicates that additional contributions, e.g., those
from initial-state interactions, may be needed to account for the observed $v_2$ signal
of prompt \cPJgy\ mesons for high-multiplicity \pPb\ events reported in this Letter.

Motivated by the possible formation of a hydrodynamically expanding
QGP medium in small systems, the elliptic flow signals for \PKzS, \PgL, \cPJgy\ and \PDz\ hadrons are compared
as a function of transverse kinetic energy (\ket) in Fig.~\ref{fig:jpsiv2resultncq} (lower),
to account for the mass difference among the four hadron species~\cite{Abelev:2007qg,Adare:2006ti}. Here,
the values of $v_{2}^\text{sub}$ and \ket\ are both divided by the number
of constituent quarks, $n_\Pq$, to represent the collective flow signal at the partonic
level in the context of the quark coalescence model~\cite{Molnar:2003ff,Greco:2003xt,Fries:2003vb},
which postulates that the elliptic flow signal of a hadron is a sum of contributions from individual constituent quark flow values.
As was previously reported in \pPb\ collisions~\cite{Khachatryan:2014jra,Sirunyan:2018toe},
a scaling of $n_\Pq$-normalized $v_{2}^\text{sub}$ values is observed
between the \PKzS\ meson and \PgL\ baryon, shown in Fig.~\ref{fig:jpsiv2resultncq} (lower).
This scaling between light baryon and meson species systems produced in the collision
(known as the number-of-constituent-quark or NCQ scaling)
was first discovered in \AonA colliding systems~\cite{Adams:2003am,Abelev:2007qg,Adare:2006ti},
indicating that collectivity is first developed among the partons,
which later recombine into final-state hadrons.
The elliptic flow signal per quark ($v_{2}^\text{sub}/n_\Pq$) for prompt \cPJgy\ mesons at
low $\ket/n_\Pq$ range is consistent with those of \PKzS, \PgL, and prompt \PDz\ hadrons within
large statistical uncertainties for the current data. There is a hint that the prompt \cPJgy\ meson data
tend to fall on the same trend as those of \PKzS\ and \PgL\ baryons, all of which are
above the prompt \PDz\ meson data.
However, the difference between the present prompt \PDz\ and \cPJgy\ meson results
deviates from 0 with a significance of only about 1.2 standard deviations at $\ket/n_\Pq \approx 0.4$\GeV.
A more definitive conclusion could be drawn with future high precision data.
For $\ket/n_\Pq > 1\GeV$, the $v_{2}^\text{sub}/n_\Pq$ for prompt \PDz\ and \cPJgy\ mesons are
consistently below that of the \PKzS\ meson.
An indication of smaller $v_{2}^\text{sub}/n_\Pq$ values for \cPJgy\ mesons than
for \PDz\ mesons is seen for $\ket/n_\Pq > 1\GeV$.
As \cPJgy\ mesons contain two charm quarks, while \PDz\ mesons contain a charm and a light-flavor quark,
this observation would be consistent with a weaker collective behavior of heavy-flavor quarks
than light quarks, possibly a consequence of
the much smaller size of the collision system. Future data with improved precision will provide
crucial insights to fully constrain the collective behavior of light- and heavy-flavor quarks
in high-multiplicity, small systems.

\section{Summary}

In summary, the elliptic flow harmonic ($v_2$) for prompt \cPJgy\ mesons in high-multiplicity
proton-lead (\pPb) collisions at $\sqrtsNN\ = 8.16\TeV$ is presented as a function of transverse momentum (\pt).
Positive $v_2$ values are observed for prompt \cPJgy\ mesons at forward rapidity
($-2.86<y_\mathrm{cm}<-1.86$ or $0.94<y_\mathrm{cm}<1.94$) over a wide \pt range ($2<\pt<8$\GeV).
This observation provides evidence for charm quark collectivity
in high-multiplicity \pPb\ collisions, similar to that first observed for light-flavor hadrons.
The observed ordering of $v_2$ among light-flavor,
open and hidden heavy-flavor hadrons at intermediate and high-\pt\ regions (e.g., above 4\GeV) adds support to
the earlier conclusion that heavy quarks exhibit weaker collective behavior than light quarks or gluons in small
systems.
For particle transverse kinetic energy per constituent quark values less than 1\GeV,
the $v_2$ of prompt \cPJgy\ mesons is consistent with prompt \PDz, \PKzS\ and \PgL\ hadrons,
within current uncertainties. A model calculation based on final-state interactions between charm quarks
and a QGP medium in \pPb\ collisions significantly underestimates the measured prompt \cPJgy\ $v_2$ signal.
The new prompt \cPJgy\ meson results,
together with previous results for light-flavor and open heavy-flavor hadrons,
provide novel insights into the dynamics of the heavy quarks produced
in small systems that lead to high final-state multiplicities.

\begin{acknowledgments}
We congratulate our colleagues in the CERN accelerator departments for the excellent performance of the LHC and thank the technical and administrative staffs at CERN and at other CMS institutes for their contributions to the success of the CMS effort. In addition, we gratefully acknowledge the computing centres and personnel of the Worldwide LHC Computing Grid for delivering so effectively the computing infrastructure essential to our analyses. Finally, we acknowledge the enduring support for the construction and operation of the LHC and the CMS detector provided by the following funding agencies: BMBWF and FWF (Austria); FNRS and FWO (Belgium); CNPq, CAPES, FAPERJ, FAPERGS, and FAPESP (Brazil); MES (Bulgaria); CERN; CAS, MoST, and NSFC (China); COLCIENCIAS (Colombia); MSES and CSF (Croatia); RPF (Cyprus); SENESCYT (Ecuador); MoER, ERC IUT, and ERDF (Estonia); Academy of Finland, MEC, and HIP (Finland); CEA and CNRS/IN2P3 (France); BMBF, DFG, and HGF (Germany); GSRT (Greece); NKFIA (Hungary); DAE and DST (India); IPM (Iran); SFI (Ireland); INFN (Italy); MSIP and NRF (Republic of Korea); MES (Latvia); LAS (Lithuania); MOE and UM (Malaysia); BUAP, CINVESTAV, CONACYT, LNS, SEP, and UASLP-FAI (Mexico); MOS (Montenegro); MBIE (New Zealand); PAEC (Pakistan); MSHE and NSC (Poland); FCT (Portugal); JINR (Dubna); MON, RosAtom, RAS, RFBR, and NRC KI (Russia); MESTD (Serbia); SEIDI, CPAN, PCTI, and FEDER (Spain); MOSTR (Sri Lanka); Swiss Funding Agencies (Switzerland); MST (Taipei); ThEPCenter, IPST, STAR, and NSTDA (Thailand); TUBITAK and TAEK (Turkey); NASU and SFFR (Ukraine); STFC (United Kingdom); DOE and NSF (USA).

\hyphenation{Rachada-pisek} Individuals have received support from the Marie-Curie programme and the European Research Council and Horizon 2020 Grant, contract No. 675440 (European Union); the Leventis Foundation; the A. P. Sloan Foundation; the Alexander von Humboldt Foundation; the Belgian Federal Science Policy Office; the Fonds pour la Formation \`a la Recherche dans l'Industrie et dans l'Agriculture (FRIA-Belgium); the Agentschap voor Innovatie door Wetenschap en Technologie (IWT-Belgium); the F.R.S.-FNRS and FWO (Belgium) under the ``Excellence of Science - EOS" - be.h project n. 30820817; the Ministry of Education, Youth and Sports (MEYS) of the Czech Republic; the Lend\"ulet (``Momentum") Programme and the J\'anos Bolyai Research Scholarship of the Hungarian Academy of Sciences, the New National Excellence Program \'UNKP, the NKFIA research grants 123842, 123959, 124845, 124850 and 125105 (Hungary); the Council of Science and Industrial Research, India; the HOMING PLUS programme of the Foundation for Polish Science, cofinanced from European Union, Regional Development Fund, the Mobility Plus programme of the Ministry of Science and Higher Education, the National Science Center (Poland), contracts Harmonia 2014/14/M/ST2/00428, Opus 2014/13/B/ST2/02543, 2014/15/B/ST2/03998, and 2015/19/B/ST2/02861, Sonata-bis 2012/07/E/ST2/01406; the National Priorities Research Program by Qatar National Research Fund; the Programa Estatal de Fomento de la Investigaci{\'o}n Cient{\'i}fica y T{\'e}cnica de Excelencia Mar\'{\i}a de Maeztu, grant MDM-2015-0509 and the Programa Severo Ochoa del Principado de Asturias; the Thalis and Aristeia programmes cofinanced by EU-ESF and the Greek NSRF; the Rachadapisek Sompot Fund for Postdoctoral Fellowship, Chulalongkorn University and the Chulalongkorn Academic into Its 2nd Century Project Advancement Project (Thailand); the Welch Foundation, contract C-1845; and the Weston Havens Foundation (USA).
\end{acknowledgments}

\bibliography{auto_generated}
\cleardoublepage \appendix\section{The CMS Collaboration \label{app:collab}}\begin{sloppypar}\hyphenpenalty=5000\widowpenalty=500\clubpenalty=5000\vskip\cmsinstskip
\textbf{Yerevan Physics Institute, Yerevan, Armenia}\\*[0pt]
A.M.~Sirunyan, A.~Tumasyan
\vskip\cmsinstskip
\textbf{Institut f\"{u}r Hochenergiephysik, Wien, Austria}\\*[0pt]
W.~Adam, F.~Ambrogi, E.~Asilar, T.~Bergauer, J.~Brandstetter, M.~Dragicevic, J.~Er\"{o}, A.~Escalante~Del~Valle, M.~Flechl, R.~Fr\"{u}hwirth\cmsAuthorMark{1}, V.M.~Ghete, J.~Hrubec, M.~Jeitler\cmsAuthorMark{1}, N.~Krammer, I.~Kr\"{a}tschmer, D.~Liko, T.~Madlener, I.~Mikulec, N.~Rad, H.~Rohringer, J.~Schieck\cmsAuthorMark{1}, R.~Sch\"{o}fbeck, M.~Spanring, D.~Spitzbart, A.~Taurok, W.~Waltenberger, J.~Wittmann, C.-E.~Wulz\cmsAuthorMark{1}, M.~Zarucki
% ...
\vskip\cmsinstskip
% ...
1:  Also at Vienna University of Technology, Vienna, Austria\\
% ...
\end{sloppypar}
\end{document}